\documentclass[aps,showpacs,superscriptaddress,twocolumn,prl]{revtex4-1}
\usepackage{amsmath}
\usepackage{amssymb}
\usepackage{bbold}
\usepackage{graphicx}
\usepackage{ulem}
\usepackage[hidelinks]{hyperref}
\usepackage{kbordermatrix}
\usepackage{xcolor}

\newcommand{\ket} [1] {\vert #1 \rangle}

\newcommand{\mean}[1]{\langle #1 \rangle}
\newcommand{\Tr}{\mathop{\mathrm{Tr}}}

\renewcommand{\kbldelim}{(}
\renewcommand{\kbrdelim}{)}

\begin{document}
\title{Bounding the sets of classical and quantum correlations in networks}

\author{Alejandro Pozas-Kerstjens}
\affiliation{ICFO-Institut de Ciencies Fotoniques, The Barcelona Institute of Science and Technology, 08860 Castelldefels (Barcelona), Spain}

\author{Rafael Rabelo}
\affiliation{Instituto de Fisica ``Gleb Wataghin'', Universidade Estadual de Campinas, CEP 13083-859, Campinas, Brazil}

\author{\L ukasz Rudnicki}
\affiliation{International Centre for Theory of Quantum Technologies (ICTQT), University of Gda{\'n}sk, 80-308, Gda{\'n}sk, Poland}
\affiliation{Max-Planck-Institut f\"ur die Physik des Lichts, Staudtstra\ss e 2, 91058 Erlangen, Germany}
\affiliation{Center for Theoretical Physics, Polish Academy of Sciences, Aleja Lotnik\'ow 32/46, 02-668 Warsaw, Poland}

\author{Rafael Chaves}
\affiliation{International Institute of Physics, Universidade Federal do Rio Grande do Norte, Campus Universitario, Lagoa Nova, Natal-RN 59078-970, Brazil}
\affiliation{School of Science and Technology, Federal University of Rio Grande do Norte, 59078-970 Natal, Brazil}

\author{Daniel Cavalcanti}
\affiliation{ICFO-Institut de Ciencies Fotoniques, The Barcelona Institute of Science and Technology, 08860 Castelldefels (Barcelona), Spain}

\author{Miguel Navascu\'es}
\affiliation{Institute for Quantum Optics and Quantum Information (IQOQI) Vienna, Austrian Academy of Sciences, Boltzmanngasse 3, 1090 Vienna, Austria}

\author{Antonio Ac\'in}
\affiliation{ICFO-Institut de Ciencies Fotoniques, The Barcelona Institute of Science and Technology, 08860 Castelldefels (Barcelona), Spain}
\affiliation{ICREA, Pg. Lluis Companys 23, 08010 Barcelona, Spain}

\begin{abstract}
We present a method that allows the study of classical and quantum correlations in networks with causally independent parties, such as the scenario underlying entanglement swapping.
By imposing relaxations of factorization constraints in a form compatible with semidefinite programming, it enables the use of the Navascu\'es-Pironio-Ac\'in hierarchy in complex quantum networks.
We first show how the technique successfully identifies correlations not attainable in the entanglement-swapping scenario.
Then we use it to show how the nonlocal power of measurements can be activated in a network: there exist measuring devices that, despite being unable to generate nonlocal correlations in the standard Bell scenario, provide a classical-quantum separation in an entanglement swapping configuration.
\end{abstract}

\maketitle

Quantum correlations are at the core of quantum information science~\cite{horodecki2009entanglement,brunner2013nonlocality}.
As proven by the violation of Bell inequalities~\cite{bell1964einstein}, quantum theory is nonlocal, in the sense that there exist correlations between outcomes of measurements performed on distant entangled quantum systems that are incompatible with any explanation involving just local hidden variables (LHV).
Quantum nonlocality is a powerful resource that grounds protocols for secure cryptography~\cite{ekert1991QKD,barrett2005qkd,acin2007qkd}, randomness certification~\cite{colbeck2007thesis,pironio2010randomness}, self-testing~\cite{mayers2004testing} or distributed computing~\cite{burhman2010complexity}.
Therefore, it is crucial to develop ways to test the incompatibility of a given correlation with LHV models, that is, to detect whether the correlation contains some nonlocal features that can be harnessed.

Nowadays, fast progress towards advanced demonstrations of quantum communication networks require to go beyond the two-party scenario and characterize networks of growing complexity, providing the tools to witness the nonclassicality of quantum correlations. 
To that aim, the framework of causal networks~\cite{pearl} and its quantum generalizations~\cite{fritz2012networks,henson2014gdag,Chaves2015a,Ried2015,Pienaar2015,Costa2016,Allen2017} have played an insightful role.
Causal networks not only allow to derive Bell's theorem from a causal inference perspective~\cite{wood2015discovery,Bellcausal} but also provide generalizations to more complex scenarios such as quantum networks with several sources~\cite{Chaves2016,branciard2010bilocality,branciard2012bilocality,fritz2012networks} or involving communication among the parties~\cite{Quantuminst1,Quantuminst2,Causalhier}.

Despite all recent advances, the understanding of the structure of correlations in networks remains very limited.
The most general method to characterize classical network correlations relies on algebraic geometry~\cite{Geiger1999} that, in practice, is limited to very simple cases.
Motivated by that, alternative methods have been proposed that either are limited to very specific networks~\cite{branciard2010bilocality,branciard2012bilocality}, or do not have a clear path for a quantum generalization~\cite{Chaves2014b,Chaves2016,wolfe2016inflation,Kela2017}.
An important recent advance is the development of the \textit{inflation technique}~\cite{wolfe2016inflation,navascues2017inflation}, that allows for the characterization of classical and general, nonsignaling network correlations.
Obtaining an analogous method to discriminate between quantum and supraquantum correlations is a subject of current research~\cite{wolfe2018quantuminflation}, but to date, the only known method for quantum correlations, the Navascu\'es-Pironio-Ac\'in (NPA) hierarchy~\cite{npashort,npalong}, applies solely to networks akin to a Bell scenario.

In this work we propose a technique to impose relaxations of factorization constraints in semidefinite programs.
When used in the framework of causality, it allows identifying correlations incompatible with causal explanations where some of the observed variables are independent.
The resulting method can be applied to any quantum network with parties that are causally independent, and can also be easily adapted to classical network correlations by incorporating extra conditions associated to measurement commutativity~\cite{baccari2017local}.
We first show that our proposal accurately identifies supraquantum correlations in the entanglement-swapping network.
Then, we use it to demonstrate that the nonlocality of measurements can be activated: we find measurement operators that, albeit unable to give rise to nonlocal correlations in a Bell scenario, can produce correlations with no classical analog when implemented in an entanglement-swapping configuration.

\paragraph{Compatibility with causal hypotheses.}
A natural tool used for reasoning about correlations in network scenarios is Bayesian networks~\cite{pearl,fritz2012networks}.
These are directed acyclic graphs (DAGs) that encode hypotheses on the causal structure underlying the correlations observed among different variables.
In Bayesian networks, nodes represent random variables and directed edges represent causal relations: the variable at the origin influences the value of the variable in the head.
The network may include hidden latent variables that could also be necessary to explain the observed correlations. 

\begin{figure}
  \centering
  \begin{tabular}{c}
    \hspace{7pt}\includegraphics[width=.41\textwidth]{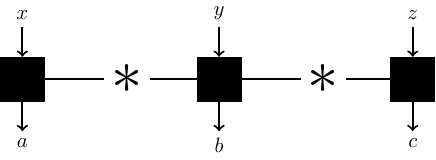}
    \\
    \includegraphics[width=.45\textwidth]{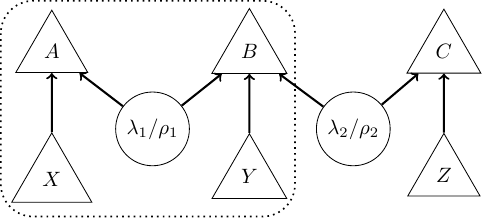}
  \end{tabular}
  \caption{The tripartite line scenario in (top) quantum information and (bottom) causal network representations. The encircled subgraph represents the DAG underlying the bipartite Bell scenario. Triangular nodes represent observable variables, while round nodes represent latent variables. Note that the extreme parties $A$ and $C$ are causally disconnected, so when the central party $B$ is omitted, correlation functions factorize.}
  \label{fig:bilocality}
\end{figure}

Typical experiments in quantum information, such as a Bell test or entanglement swapping, can easily be represented in this language.
Sources preparing states, which are not directly empirically observable, are represented by latent variables, while measurement choices and outputs define the observed variables, as represented in Fig.~\ref{fig:bilocality}.
A network provides a classical-quantum separation whenever there exist quantum sources producing correlations among the observed variables that are impossible to attain by classical means in the same network.
The scenario underlying a Bell test is an example of such a network, but it is not the only one~\cite{fritz2012networks,henson2014gdag,pienaar2017interesting,branciard2010bilocality,branciard2012bilocality,Quantuminst1,Quantuminst2}.

Assessing whether a given correlation is compatible with a causal network with latent variables is hard.
Even though in the classical case some general recipes are known~\cite{Geiger1999,Chaves2014b,Chaves2016,wolfe2016inflation,Kela2017}, they cannot be easily generalized for quantum correlations.
For instance, while the inflation technique~\cite{wolfe2016inflation,navascues2017inflation} can identify if a correlation is non-classical, it cannot discern whether its origin is quantum or supraquantum.
Only for Bell-type networks, where measurements are applied on different shares of a single multipartite quantum state, can the NPA hierarchy~\cite{npashort,npalong} be used to discard a quantum origin of the correlations.

The NPA hierarchy works under the following observation: assume a multipartite probability distribution $p(a,b,\dots|x,y,\dots)$ has a quantum realization, so there exists a quantum state $\rho$ and measurement operators $\{\Pi^q_{o|i}\}_{i,o}$ for $q\,{=}\,\textsc{a},\textsc{b}\dots$ such that \mbox{$p(a,b,\dots|x,y,\dots)\,{=}\,\Tr(\rho\Pi^\textsc{a}_{a|x}\otimes\Pi^\textsc{b}_{b|y}\otimes\dots)$}.
For every set of products of any size of the measurement operators, $\mathcal{S}\,{=}\,\{S_1,\dots,S_n\}$, the moment matrix $\Gamma$ whose matrix elements are given by
\begin{equation}
    \Gamma_{i,j}=\Tr\left(\rho S_i^\dagger S_j\right)\qquad S_i, S_j\in\mathcal{S}
    \label{eq:gamma}
\end{equation}
is positive semidefinite.
To test whether a given probability distribution admits a quantum realization, one chooses a set $\mathcal{S}$, builds a moment matrix of the form~\eqref{eq:gamma}, where some entries will be computable from the distribution while the rest remain as variables, and looks for a variable assignment that makes it positive semidefinite.
Finding a set $\mathcal{S}$ for which the associated $\Gamma$ cannot be made positive semidefinite signals the distribution as not compatible with a quantum origin.
This type of reasoning also leads to a well-known tool for entanglement detection in continuous variables~\cite{Vogel2005}.

In the case of multipartite Bell scenarios several linear constraints between the matrix entries of $\Gamma$ arise due to the properties of quantum measurements, so one can use semidefinite programming to check the existence of a variable assignment that makes $\Gamma$ positive semidefinite.
This is not the case in general multipartite scenarios with multiple sources (states) because the network conditions may give rise to nonlinear constraints that cannot be imposed to the semidefinite programs.

For illustration, consider the line scenario, a generalization of the Bell scenario underlying quantum repeaters where parties are arranged in a line and two consecutive parties are linked together by a latent variable.
The simplest of these instances corresponds to an entanglement swapping experiment involving three parties, as represented in Fig.~\ref{fig:bilocality}.
Correlations compatible with this scenario are such that the marginal distributions resulting after discarding any non-extremal party factorize,
\begin{align}\label{eq:factor}
    &\sum_{o_k}\,p(o_1,\dots, o_k,\dots, o_n|i_1,\dots, i_k,\dots, i_n) \\
        &=\!p(o_1,\dots, o_{k-1}|i_1,\dots\!, i_{k-1})p(o_{k+1},\dots, o_{n}|i_{k+1},\dots, i_{n}),\notag
\end{align}
where $i_k$ and $o_k$ denote the input and output variable for party $k$.
While illustrated here for the line, these independence constraints apply to many networks.
They are nonlinear and nonconvex, and thus cannot be imposed in semidefinite programs.
In the following we propose a method to partially overcome this limitation by imposing semidefinite-program-compatible relaxations of conditions of the type of Eq.~\eqref{eq:factor}.

\paragraph{Convex relaxation of independence relations.}
In order to deal with independence relations in semidefinite programming, we propose performing a \textit{scalar extension} of the moment matrices involved in the NPA hierarchy: given a set of products of measurement operators, $\mathcal{S}$, which produces a specific moment matrix $\Gamma$, we complement it with extra operators of the form $S_i\mean{S_j}$, $S_i\mean{S_j}\mean{S_k},\dots$, where $S_i$, $S_j$, $S_k,\dots$ are products of operators not necessarily belonging to the original set $\mathcal{S}$.

The additional generating operators give rise to matrix elements that represent factorized quantities, which can then be related via equality constraints to elements in the original moment matrix that should factorize. 
In the same spirit as the original NPA hierarchy, some of the matrix elements in $\tilde{\Gamma}$ can be computed from the given probability distribution, the rest remaining as variables, and if the probability distribution is compatible with the causal scenario then there exists a variable assignment such that $\tilde{\Gamma}\,{\succeq}\, 0$.

Since $\Gamma$ is a principal submatrix of $\tilde{\Gamma}$, the positivity of the latter implies the positivity of the former.
If we cannot find a positive semidefinite completion of $\tilde{\Gamma}$, it follows that the correlation under scrutiny is not compatible with the proposed causal explanation.
If, furthermore, we cannot find a non-negative completion for $\Gamma$, the correlation cannot be generated even by conducting measurements on a global quantum state.

To illustrate the method let us consider the tripartite line scenario of Fig.~\ref{fig:bilocality}.
Because of the network geometry, all entries in the moment matrix generated by operator strings that only contain operators of the extreme parties $A$ and $C$ factorize.
Consider the moment matrix generated by the extended set of operators $\{\mathbb{1}, A_0A_1, C_0C_1, \mean{A_0A_1}\mathbb{1}\}$,
\begin{equation}
\tilde{\Gamma} = \kbordermatrix{
                        & \mathbb{1} & A_0A_1 & C_0C_1 & \mean{A_0A_1}\mathbb{1} \\
\mathbb{1}              & 1          & v_1    & v_2    & v_3                     \\
(A_0A_1)^{\dagger}                  &            & 1      & v_4    & v_5                     \\
(C_0C_1)^{\dagger}                  &            &        & 1      & v_6                     \\
\mean{A_0A_1}^{*}\mathbb{1} &            &        &        & v_7
},
\label{emm}
\end{equation}
where the lower triangle has been omitted since $\tilde{\Gamma}$ is Hermitian and we assume that every operator satisfies $O^\dagger O\,{=}\,\mathbb{1}$.
The insertion of the extra column labelled by the operator $\mean{A_0A_1}\mathbb{1}$, which is an operator equal to the identity times the unknown scalar factor $\mean{A_0A_1}$, gives rise to a number of equality constraints that relate elements in the top-left $3\,{\times}\,3$ submatrix (which is the corresponding $\Gamma$) with elements in the last column.
These identifications are $v_1\,{=}\,v_3$, $v_5\,{=}\,v_7$, and $v_4\,{=}\,v_6^{*}$.
The latter is actually that which imposes the independence constraint \mbox{$\mean{A_0A_1C_0C_1}\,{=}\,\mean{A_0A_1}\mean{C_0C_1}$}, a causal constraint implied by the tripartite-line structure.

We note that additional constraints can be imposed in $\tilde{\Gamma}$, namely $v_4\,{=}\,v_1^*v_2$ (also arising from the causal structure) and $v_5\,{=}\,|v_1|^2$. These constraints are nonlinear and will not be enforced in our method, since we wish to keep it solvable via semidefinite programming. Furthermore, as shown below, they are not necessarily required for providing useful information about correlations in networks.

The independence constraints exploited in scalar extension apply both to classical and quantum networks.
In the classical case it is possible to incorporate additional linear constraints to the moment matrix that account for the fact that in classical theory all measurement operators, including those performed by the same party, commute.
This creates a hierarchy over relaxations of the set of classical correlations, which has been used in the context of entanglement detection in Ref.~\cite{baccari2017local}.
Therefore, the method can be employed to study both classical and quantum correlations in networks.

\paragraph{Identifying supraquantum correlations.}
Simple scalar extensions allow recovering the results of Ref.~\cite{branciard2012bilocality}. The tripartite line scenario of Fig.~\ref{fig:bilocality} underlies the entanglement swapping protocol.
Being the simplest quantum network beyond Bell, it has received considerable attention~\cite{branciard2010bilocality,branciard2012bilocality,Chaves2016}.
Non-linear Bell-like inequalities have been obtained to discern in a device-independent manner whether correlations observed in tripartite scenarios are compatible or not with \textit{bilocal hidden variable models} of the form
\begin{align}
    p(a,b,c|x,y,z)=&\int_{\Lambda_1}\text{d}\lambda_1\int_{\Lambda_2}\text{d}\lambda_2\,
    p(\lambda_1)p(\lambda_2) \notag\\
    &\times p(a|x,\lambda_1)\,p(b|y,\lambda_1,\lambda_2)\,p(c|z,\lambda_2).
    \label{eq:bilocal}
\end{align}

In fact, similar queries can be done on whether a particular distribution can have a \textit{biquantum model}, i.e., if it can be written in the form
\begin{equation}
    p(a,b,c|x,y,z)\!=\!\Tr\left(\Pi_{a|x}^\textsc{a}\otimes\Pi_{b|y}^\textsc{b${}_1$b${}_2$}\otimes\Pi_{c|z}^\textsc{c} \rho_\textsc{ab${}_1$}\otimes\rho_\textsc{b${}_2$c}\right).
    \label{eq:biquantum}
\end{equation}

As an example of use of the scalar extension technique, we recover two known results about the entanglement swapping configuration where each party has a binary input and output, $x,y,z,a,b,c\,{=}\,0,1$.
Consider the one-parameter family of correlations
\begin{equation*}
    P_v=vP^{22}+(1-v)P_0,
\end{equation*}
where $0\,{\leq}\, v\,{\leq}\, 1$, $P_0$ is a noise term \mbox{$P_0(a,b,c|x,y,z)\,{=}\,1/8$} for all inputs and outputs, and $P^{22}$ is~\cite{branciard2012bilocality}
\begin{equation*}
    P^{22}(a,b,c|x,y,z)=\frac{1}{8}\left[1+\left(-1\right)^{a+b+c+xy+yz}\right].
\end{equation*}

This distribution fails to have a biquantum model for any $v\,{>}\,1/2$, and a bilocal hidden variable model for any $v\,{>}\,1/4$.
Using the scalar extension construction we are able to reproduce these results.
The explicit calculations performed in this section are shown in the Computational Appendix, which can be found in~\cite{compapp}.

For comparing against biquantum models, we consider an extension of the NPA level $\mathcal{S}_3$~\cite{npalong} with a minimal set of operators needed to impose constraints on all elements in the moment matrix that should factorize.
This minimal set has the form $\{\mean{S_1}\mathbb{1},\dots,\mean{S_n}\mathbb{1}\}$, where $S_i$ is any combination of operators of party $A$ of length $2\,{\leq}\, l \,{\leq}\, 5$.
We generate the necessary moment matrices $\tilde{\Gamma}$ with \textsc{Ncpol2sdpa}~\cite{ncpol2sdpa}, and optimize their smallest eigenvalues using \textsc{Mosek}~\cite{mosek}, \textsc{SeDuMi}~\cite{sedumi} and SDPT3~\cite{sdpt3}.
For visibilities $v\,{>}\,1/2$ the optimal smallest eigenvalues found are negative~\footnote{We adopt as stopping criterion for the solvers that the absolute gap between the primal and dual objectives is smaller than $10^{-12}$. We consider the optimal smallest eigenvalue to be negative whenever both the primal and dual objectives are negative. We accept the output of an optimization problem as valid whenever the results given by all solvers coincide up to a relative error of $10^{-6}$.}. This renders the family of distributions $P_{v>1/2}$ incompatible with biquantum models.
For $v\,{=}\,1/2$ the distribution is equal to $P_{1/2}\,{=}\,P_Q^{22}$, defined in Ref.~\cite{branciard2012bilocality}, which is known to have a quantum realization.

In order to compare $P_v$ against bilocal models, we impose the additional constraint that operators representing different measurements of a same party commute~\cite{baccari2017local}.
Using the corresponding generating set $\mathcal{S}_3$, in this case an extension with the single operator $\mean{A_0A_1}\mathbb{1}$ suffices to discard the existence of a bilocal model of the correlations whenever the visibility satisfies $v\,{>}\,1/4$.
This is the same result than that shown in Ref.~\cite{branciard2012bilocality} for noisy versions of the quantum distribution $P_Q^{22}(V)$, since $P_{v=1/4}\,{=}\,P_Q^{22}(V\,{=}\,1/2)$.
When $v \,{\leq}\, 1/4$, the distribution is known to have a bilocal model.

\paragraph{Nonlocality of measurement activation.}
In the following we use scalar extension to demonstrate how the nonlocal power of measurement devices can be activated in a network structure: measurements that do not lead to nonlocal correlations in the standard Bell scenario, do it when arranged in a network.
Similar effects are known for quantum states~\cite{cavalcanti2011network}.

Consider a scenario in which one has access to three measuring devices. One implements a single four-output measurement.
The remaining two devices each implement two measurements of binary outputs with limited detection efficiency: for all measurements there is a probability $1\,{-}\,\eta$ that a third (lossy) outcome is observed.
We denote by $\eta_1$ and $\eta_2$ these efficiencies.
What are their values so that nonlocal correlations can be certified with these three devices acting on a quantum state?
Since one of the devices implements only one measurement, the only possibility is to run a standard bipartite Bell test with the other two, possibly conditioned on one of the measurement outputs of the first.
Critical values for the detection efficiencies such that no nonlocal correlations can be observed with two-output measurements can be obtained from Ref.~\cite{massar2003detection}: in case one device is perfect, say $\eta_1\,{=}\,1$, it is impossible to observe nonlocal correlations whenever $\eta_2\,{<}\,1/2$; if both devices have the same efficiency, a local model for the correlations always exists if $\eta_1\,{=}\,\eta_2\,{\leq}\, 2/3$.

We now arrange these devices in a tripartite-line scenario and make use of scalar extension to determine detection efficiencies for which nonclassical correlations can be observed in the network.
In particular, we focus on the case where the sources send partially entangled states $\ket{\theta_{ij}}\,{=}\,\cos\theta_{ij}\ket{00}\,{+}\,\sin\theta_{ij}\ket{11}$, the measurement device of party $B$ makes a perfect, four-outcome Bell state measurement in the standard basis $\{\ket{\phi^+},\ket{\phi^-},\ket{\psi^+},\ket{\psi^-}\}$, and the measurements performed by parties $A$ and $C$ have the form~\cite{bancal2014eberhard}:
\begin{align*}
    A_0 &= \cos\alpha_0 Z - \sin\alpha_0 X,\qquad C_0 = \cos\alpha_0 Z + \sin\alpha_0 X, \\
    A_1 &= \cos\alpha_1 Z + \sin\alpha_1 X,\qquad C_1 = \cos\alpha_1 Z - \sin\alpha_1 X.
\end{align*}

With probability $1\,{-}\,\eta_i$, party $i\in\{A,C\}$ produces a third, ``measurement failed'' result.
We are interested in knowing how inefficient the measurement devices of the extreme parties $A$ and $C$ can be and still be able to produce correlations that cannot be explained by a bilocal hidden variable model of the type~\eqref{eq:bilocal}.
The results are shown in Table~\ref{table:results}, where we also vary the entanglement of the prepared pure states.
All results have been achieved with the generating operator set corresponding to the (commuting) NPA level $\mathcal{S}_3$, extended with the set of four operators $\{\mean{\Pi^\textsc{c}_{c_1|0}\Pi^\textsc{c}_{c_2|1}}\mathbb{1}:\,c_1, c_2\in\{0,1\}\}$, where $\Pi^\textsc{c}_{o|i}$ is the projector on the outcome $o$ of the $i$th measurement of party $C$.

\begin{table}
\begin{tabular}{l|c|c|c}
    Assumptions                          & $\eta_\text{min}$ & $\theta_\text{min}$ & $(\eta,\theta)_\text{ex}$ \\
    \hline\hline
    \begin{tabular}{ll} $\eta_\textsc{a}=1$, & \hspace{4pt}$\theta_\textsc{ab}=\pi/4$ \end{tabular}        & $<10^{-5}$       & $<10^{-4}$         & (0.0001, 0.1250)      \\
    \begin{tabular}{ll} $\eta_\textsc{a}=1$, & \hspace{4pt}$\theta_\textsc{ab}=\theta_\textsc{bc}$ \end{tabular}      & $<10^{-5}$      & $<10^{-4}$         & (0.0444, 0.1000)      \\
    \begin{tabular}{ll} $\eta_\textsc{a}=\eta_\textsc{c}$, & $\theta_\textsc{ab}=\pi/4$ \end{tabular}    & 0.6085       & 0.0010         & (0.6389, 0.6545)      \\
    \begin{tabular}{ll} $\eta_\textsc{a}=\eta_\textsc{c}$, & $\theta_\textsc{ab}=\theta_\textsc{bc}$ \end{tabular} & 0.5291      & 0.0070         & (0.5626, 0.1751)      
\end{tabular}
\caption{Upper bounds to the smallest detection efficiency and entanglement needed to generate nonbilocal correlations in the tripartite-line scenario. The values on the second and third columns $(\eta_\text{min},\theta_\text{min})$ are not related, so in general one needs $\theta\,{>}\,\theta_\text{min}$ to be able to discard bilocal models with $\eta\,{=}\,\eta_\text{min}$ and vice-versa. The last column shows an example of small combined values of the parameters for which non-bilocal correlations can be generated.}
    \label{table:results}
\end{table}

When party $A$'s measurement device is perfect (the case in the two first rows of Table~\ref{table:results}), the observed three-party correlations do not have bilocal models even in cases when party $C$ detects as few as $0.001\%$ of all the particles that receives.
This value is well below $50\%$, the critical value needed for certifying standard nonlocality.
In fact, we believe that nonclassical correlations can be observed for all $\eta_\textsc{c}\,{>}\,0$ and the obtained critical value is a consequence of numerical issues when the detection efficiency is very low.
We also observe that very low entanglement is needed to create nonbilocal correlations.
This last finding is similar to results known for bipartite Bell scenarios~\cite{gisin1991anyentangled}.

In the case of two inefficient devices, if we fix the state prepared by the source between parties $A$ and $B$ to be maximally entangled, nonbilocal correlations can be established for any detection efficiency higher than $\eta_\text{min}\,{=}\,0.6085$.
Nonbilocal correlations at $\eta_\text{min}$ are generated when $\theta_\textsc{bc}\,{=}\,\pi / 4$ (so the source between parties $B$ and $C$ also distributes maximally entangled states), and the measurements performed are $A_x \,{=}\, [Z-(-1)^x X]/\sqrt{2}$, $C_z \,{=}\, [Z+(-1)^z X]/\sqrt{2}$.
For higher detection efficiencies, nonbilocal correlations can be generated with decreasing amounts of entanglement and measurements whose aperture (the relative angle $|\alpha_1 - \alpha_0|$) decreases, suggesting that this is the optimal configuration for detecting nonbilocality in a loss-resistant manner.

Varying also the entanglement in $\theta_\textsc{ab}$, correlations without bilocal models can be certified for efficiencies above \mbox{$\eta_\text{min}\,{=}\,0.5291$}, again well below $2/3$, the value necessary for certifying standard Bell nonlocality.
It is then a relevant question to ask whether the value of $\eta_\text{min}$ can be further lowered by adding more copies of the source either in an $n$-partite line or an $n$-star scenario, which we leave for future work.

\paragraph{Conclusions.}
Technological advances have been allowing us to distribute information encoded in quantum systems in increasingly complex network structures in the recent years.
With the aim of testing the hypothesis that an observed multipartite correlation has been generated in a particular quantum network, we have introduced a \textit{scalar extension} of the NPA hierarchy that imposes relaxations of the independence constraints between causally independent nodes in a network scenario.
The method can also be applied to classical correlations by introducing some extra constraints associated to the commutativity of all measurements, including those of the same observer.
We have applied this method to the study of correlations in the tripartite-line scenario, showing that correlations not explainable by bilocal hidden variable models can be established between the parties even in the case of very low detection efficiencies---at least as low as $0.001\%$.
The results imply that it is possible to activate the nonlocal properties of measurement devices by arranging them in a network geometry.
This has a clear impact on the deployment of state-of-the-art and near-future experimental quantum networks, as it lowers the requirements for certification of non-classical correlations.

It is a natural open problem to study whether the obtained hierarchy converges.
It is also relevant to analyze how these results could be used to construct quantum information protocols.
In particular, understanding how to extend the results to line scenarios with more nodes could be of relevance for the observation of non-classical correlations in repeater networks.
A similar extension can be proposed for the entanglement criteria based on various moments of continuous variables' quadratures~\cite{Vogel2005}, possibly leading to novel methods of multipartite entanglement detection.

In its core, scalar extension imposes factorization constraints in semidefinite programs, and its application to correlations in networks comes at a later stage.
Semidefinite programming for polynomial optimization is ubiquitous in many fields of research and engineering, and, thus, the application of scalar extension in other fields represents a promising new avenue of research.

\paragraph{Acknowledgments.}
The authors are grateful to Elie Wolfe, Peter Wittek, Flavio Baccari and Marco T\'ulio Quintino for insightful discussions.
This work is supported by the ERC CoG QITBOX, the AXA Chair in Quantum Information Science, Fundacio Obra Social ``la Caixa'' (LCF/BQ/ES15/10360001), the Spanish MINECO (QIBEQI FIS2016-80773-P and Severo Ochoa SEV-2015-0522), Fundacio Cellex, Generalitat de Catalunya (CERCA Program and SGR1381), the John Templeton Foundation via the Grant Q-CAUSAL No. 61084, the Serrapilheira Institute (Grant No. Serra-1708-15763), the S\~{a}o Paulo Research Foundation FAPESP (Grant No. \#2018/07258-7), the Brazilian National Council for Scientific and Technological Development (CNPq) via the National Institute for Science and Technology on Quantum Information (INCT-IQ) and Grants No. 307172/2017-1 and No. 406574/2018-9, the Brazilian agencies MCTIC and MEC, the Austrian Science fund (FWF) standalone Project No. P 30947, and the Foundation for Polish Science (IRAP project, ICTQT, Contract No. 2018/MAB/5, cofinanced by EU within Smart Growth Operational Programme).

\bibliography{bibliography}

\end{document}